\documentclass[aps,prb,twocolumn,epsfig,amsmath,showpacs]{revtex4-1}
\usepackage{epsfig}
\usepackage{dcolumn}
\usepackage{bm}
\usepackage{color}
\begin{document}
\title{Density functional study of orbital-selective magnetism in FeAs-based superconductors}

\author{Hyungju Oh}
\author{Donghan Shin}
\author{Hyoung Joon Choi}
\email[Email:\ ]{h.j.choi@yonsei.ac.kr}
\affiliation{Department of Physics and IPAP, Yonsei University, 
Seoul 120-749, Korea}

\date{\today}

\begin{abstract}
We performed spin-polarized density functional calculations of 
lanthanide-series (Ln) iron oxypnictides LnFeAsO (Ln=La, Ce, Pr, Nd, Sm, 
and Gd) with constrained Fe magnetic moments, finding that in-plane 
$d_{xy}$ and out-of-plane $d_{yz}$ orbital characters are preferred for 
small Fe magnetic moments. Comparison of LnFeAsO compounds shows that the 
antiferromagnetism (AFM) from the Fe $d_{xy}$ orbital is itinerantly driven 
by orbital-dependent Fermi-surface nesting while AFM from the Fe $d_{yz}$ 
orbital is driven by superexchange mechanism. The Fe magnetic moments of 
the two orbital characters show different coupling strengths to Fermi-surface 
electrons orbital-selectively, suggesting that they may play different roles 
in superconductivity and in AFM, and making $d$ orbital characters of the 
magnetic moment resolvable by measuring the electronic structures. 
\end{abstract}


\maketitle

LaFeAsO$_{1-x}$F$_x$ (Ref.~\onlinecite{kamihara1}) and related compounds show 
unconventional superconductivity (SC) in the vicinity of
antiferromagnetism (AFM).\cite{chenxh,cruz,zhao1,yin,mazin,yildirim,
chengf,qiu,zhao2,ren1,ren2,miyazawa,kamihara2,wang,moon}
Among various families of iron pnictides and chalcogenides,
lanthanide-series (Ln) iron oxypnictides LnFeAsO show the 
highest superconducting transition temperature ($T_c$) with doping.
Reported $T_c$ in doped LnFeAsO increases dramatically from 
26~K up to 55~K with Nd or Sm substitution for La and then 
$T_c$ decreases slightly in doped 
GdFeAsO.\cite{kamihara1,chenxh,chengf,ren1,ren2,miyazawa,kamihara2,wang}
With this strong variation of $T_c$, LnFeAsO is suited for studying 
material dependence of $T_c$.\cite{kuroki} Since FeAs-based 
materials are featured with multiple Fermi surfaces (FSs) with 
strong orbital characters, many theoretical and experimental studies 
have been done on FS nesting, local-moment interactions, and orbital 
orderings.\cite{kuroki,lebegue,lu,singh,haule,moon_2008,lee,kruger,moon_2009,chencc,lv,yi}

So far, study of orbital physics in FeAs-based materials is focused mainly
on $d_{yz}$ versus $d_{zx}$ orbital 
characters.\cite{lee,kruger,lv2,yin2,daghofer} It was 
claimed theoretically that Fe $d_{yz}$ orbital is less occupied but more 
spin-polarized than Fe $d_{zx}$ orbital, resulting in ferro-orbital 
order.\cite{lee,kruger,lv2} It was also claimed theoretically that the 
low-energy orbital polarization between $d_{yz}$ and $d_{zx}$ orbitals 
leads to the anisotropy of the optical conductivity.\cite{yin2} 
The orbital magnetization was reported to be far stronger for the $d_{yz}$ 
orbital than it is for $d_{zx}$, which suggests that the orbital degree of 
freedom strongly couples to the magnetic order.\cite{daghofer} 
All these works are focused on comparing Fe $d_{yz}$ and $d_{zx}$ orbitals
and indicate that $d_{yz}$ orbital has more important roles in magnetism or 
orbital order than $d_{zx}$ orbital does, not considering any possible role
of the other $d$ orbitals. However, it was reported, for example, that 
Fe $d_{xy}$ orbital contributes more to Fe magnetic moment than $d_{zx}$ 
orbital does.\cite{yin3} Thus, full analysis of orbital characters including 
$d_{xy}$ orbital is needed for electronic and magnetic 
properties of FeAs-based materials.

In this paper, we report spin-polarized density functional calculations of 
LnFeAsO (Ln=La, Ce, Pr, Nd, Sm, and Gd), constraining the magnitude and
the $d$ orbital character of Fe magnetic moment. We find that in-plane 
$d_{xy}$ and out-of-plane $d_{yz}$ orbital characters are energetically 
preferred for Fe magnetic moments when the moments are small and ordered 
antiferromagnetically along the $x$ direction and ferromagnetically along 
the $y$ direction. Comparison of atomic and electronic structures of 
LnFeAsO compounds shows that AFM from the Fe $d_{xy}$ orbital is 
itinerantly driven by orbital-dependent FS nesting while AFM from 
the Fe $d_{yz}$ orbital is driven by superexchange mechanism. 
The Fe magnetic moments of the two orbital characters show different coupling 
strengths to FS electrons orbital-selectively. This suggests that Fe moments
of different orbital character can be resolved by measuring the electronic 
structures and may have different role in SC and AFM.

Our calculations are based on {\em ab-initio} norm-conserving 
pseudopotentials and the Perdew-Burke-Ernzerhof-type generalized gradient 
approximation to the density functional theory (DFT), as implemented in the 
SIESTA code.\cite{sanchez} Electronic wavefunctions are expanded with 
pseudoatomic orbitals (double-$\zeta$ polarization), which yields orbital 
characters in electronic and magnetic properties straightforwardly. In our 
present work, the size of the magnetic moment ($m_\text{Fe}$) of a Fe atom 
is defined as the difference of spin-up and spin-down electrons occupying 
pseudoatomic orbitals of the Fe atom. During self-consistent iterations, 
the total-energy functional is minimized with constraints imposed on the 
electron spin density in order to control the size and the orbital character 
of the Fe magnetic moment. Since experimentally reported values of the Fe 
magnetic moment in LnFeAsO are quite smaller than the unconstrained DFT 
values of the Fe magnetic moment, we consider constraining the size of the 
Fe magnetic moment to a quite smaller value than an unconstrained one. 
To consider a Fe magnetic moment of a specific $d$-orbital character, 
constraints are imposed so that only a specified $d$ orbital can have 
different occupations of spin-up and spin-down electrons while all the 
other $d$ orbitals are spin-unpolarized. We assume an oxidation 
state of +3 for all considered Ln elements, treating their 4$f$ orbitals 
as core orbitals.\cite{nekrasov,pourovskii} Experimental lattice constants 
and atomic positions at the high-temperature tetragonal 
phase \cite{cruz,zhao1,qiu,zhao2,wang,martinelli,structure_optimization} 
are used in our calculations in order to focus on effects of Fe magnetic 
moments on the electronic and magnetic properties \cite{T_vs_O} although 
LnFeAsO is orthorhombic at low temperature. A dense {\bf k}-point grid of 
32$\times$32$\times$32 is used to determine the Fermi energy ($E_F$) and 
FS properties precisely. Spin-orbit coupling is not considered, since our 
separate test calculations show that spin-orbit coupling has only minor 
effects on properties under consideration in our present work.


\begin{figure} 
\epsfig{file=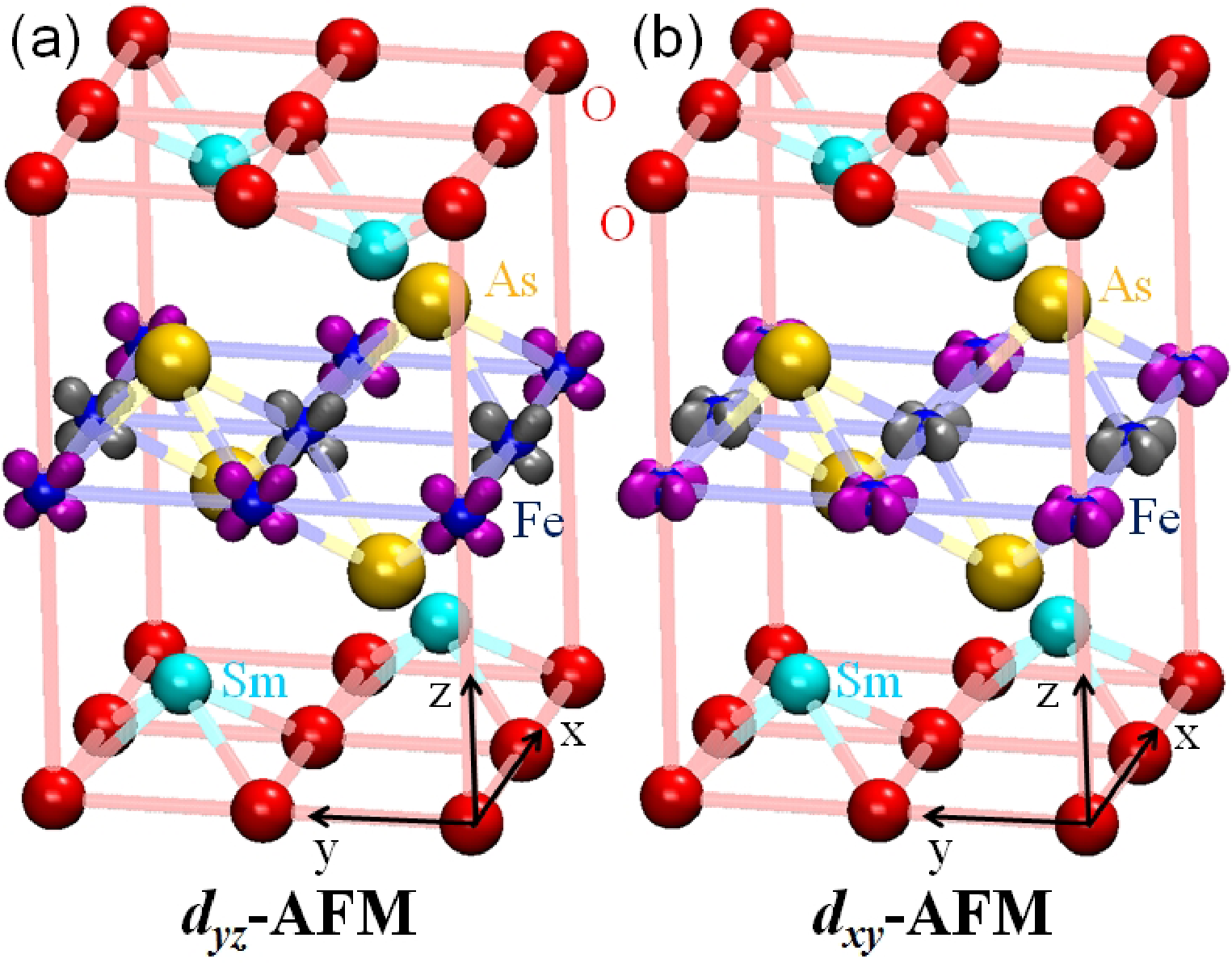,width=8.5cm,clip=}
\caption{(Color online) Orbital characters of Fe magnetic moments in 
SmFeAsO: (a) $d_{yz}$-AFM where Fe magnetic moments have only $d_{yz}$ 
orbital character and (b) $d_{xy}$-AFM where Fe magnetic moments have 
only $d_{xy}$ orbital character. In (a) and (b), the size of each Fe 
magnetic moment ($m_\text{Fe}$) is constrained to 0.10~$\mu_\text{B}$ 
as an example of small $m_\text{Fe}$. Each atomic structure is a 
$\sqrt{2}\times\sqrt{2}\times1$ supercell with four Fe atoms in the 
single-stripe-type AFM. Fe magnetic moments, in dark and light grays 
(purple and light gray) which represent opposite directions of the 
moments, are ordered antiferromagnetically along the $x$ axis and 
ferromagnetically along the $y$ axis. In (a) and (b), the $d$ orbital 
shapes of Fe magnetic moments contain 50\% of the total spin density. 
The example value of $m_\text{Fe}$ of 0.10~$\mu_\text{B}$ is about 30\% 
of the experimental value of 0.34~$\mu_\text{B}$ in SmFeAsO.\cite{kamihara2}
}
\end{figure}

An intriguing finding in our present calculations is distinctive 
$d$-orbital characters of the Fe magnetic moment in LnFeAsO in small 
$m_\text{Fe}$ regime. Figure~1 shows two prototypical cases of SmFeAsO in 
the single-stripe-type AFM with antiferromagnetic ordering along the $x$ 
direction and ferromagnetic ordering along the $y$ direction. 
When $m_\text{Fe}$ is set to 0.10 Bohr magneton ($\mu_\text{B}$) per Fe atom
as an example of small $m_\text{Fe}$,
SmFeAsO has two almost degenerate ground states. In one case [Fig.~1(a)] 
the Fe magnetic moment has out-of-plane $d_{yz}$ orbital character
while in the other case [Fig.~1(b)] the Fe magnetic moment has in-plane 
$d_{xy}$ orbital character. In the former case the Fe magnetic moment is 
made by spin polarization of electrons in $d_{yz}$ orbitals while in the 
latter case the Fe magnetic moment is made by spin polarization of electrons 
in $d_{xy}$ orbitals. When we considered Fe magnetic moments having 
$d$-orbital characters other than $d_{yz}$ and $d_{xy}$, we obtained the 
total energy which is higher than the nonmagnetic case, so we consider only 
$d_{yz}$ and $d_{xy}$ orbital characters for the Fe magnetic moment in our 
present work. In previous studies, the main concern was not
$d_{yz}$ versus $d_{xy}$ orbitals, but $d_{yz}$ versus $d_{zx}$ orbitals 
in regard to symmetry lowering from tetragonal to orthorhombic structures. 
In our present work, symmetry is lowered by imposing the single-stripe-type 
AFM and then either $d_{yz}$ or $d_{xy}$ orbital character is found
predominant in the Fe magnetic moment, interestingly.


\begin{figure} 
\epsfig{file=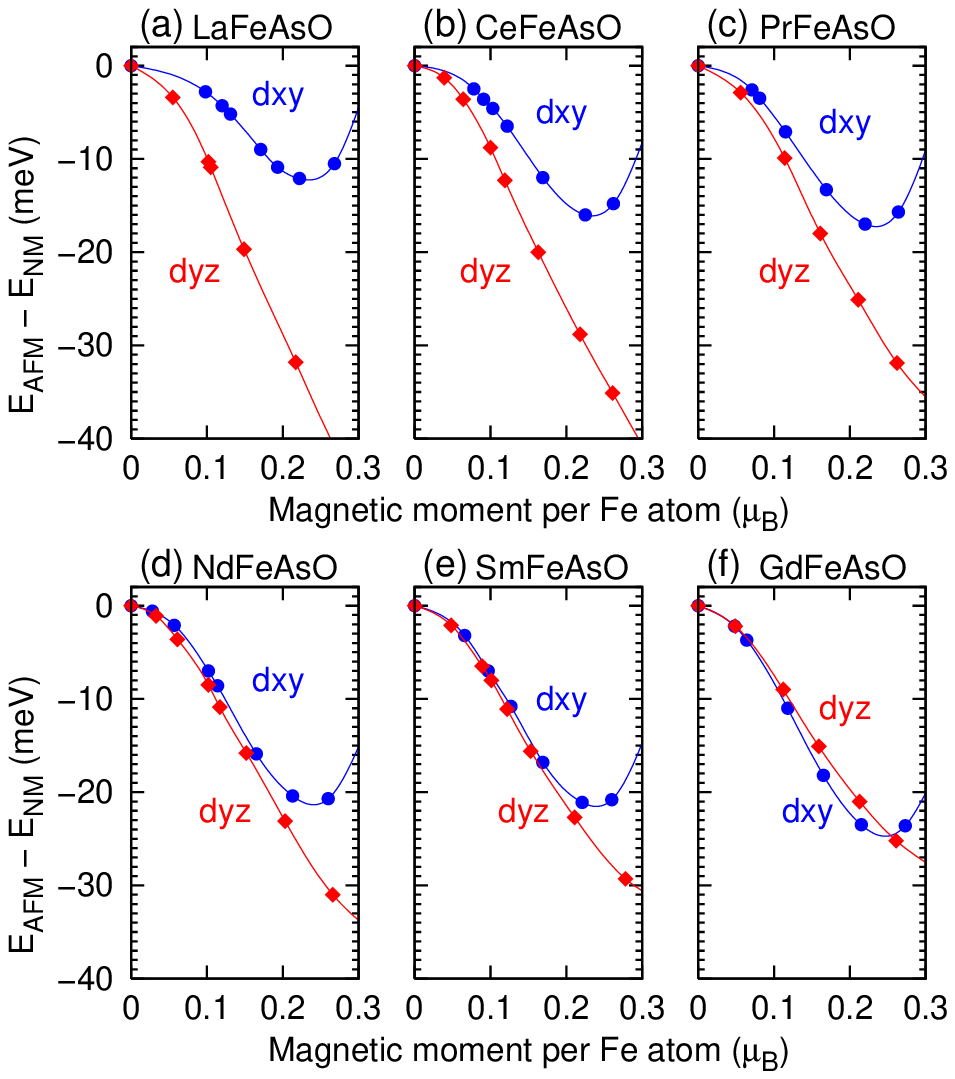,width=8.5cm,angle=0,clip=} 
\caption{(Color online) 
The total energy of LnFeAsO in $d_{xy}$- and $d_{yz}$-AFMs with constrained
Fe magnetic moments: (a) LaFeAsO, (b) CeFeAsO, (c) PrFeAsO, (d) NdFeAsO,
(e) SmFeAsO, and (f) GdFeAsO. The total energy per supercell containing 
four Fe atoms is plotted as a function of the magnetic moment per
Fe atom. Fe magnetic moments are constrained to specific values
and to either $d_{xy}$ or $d_{yz}$ orbital character.
In each plot, the total energy of the nonmagnetic phase is set to zero.
}
\end{figure}

For systematic study, we calculated the total energy of LnFeAsO in the 
single-stripe-type AFM with Fe magnetic moments constrained to have either 
$d_{yz}$ or $d_{xy}$ orbital character (which we denote as $d_{yz}$-AFM and 
$d_{xy}$-AFM, respectively) as a function of $m_\text{Fe}$, as shown 
in Fig.~2. With Ln ranging from La to Gd, we found that $d_{yz}$-AFM becomes 
less and less stable while $d_{xy}$-AFM becomes more and more stable. 
In LaFeAsO, $d_{yz}$-AFM is lower in energy than 
$d_{xy}$-AFM with the total-energy difference of about 20~meV per four Fe 
atoms when $m_\text{Fe}$ is 0.20~$\mu_\text{B}$, as shown in Fig.~2(a).
However, in the case of SmFeAsO, $d_{xy}$- and $d_{yz}$-AFMs are nearly 
degenerate when $m_\text{Fe}$ is less than 0.20~$\mu_\text{B}$, as shown
in Fig.~2(e). In GdFeAsO, $d_{xy}$-AFM has a lower total energy than 
$d_{yz}$-AFM, so $d_{xy}$-AFM becomes dominant [Fig.~2(f)].

Our additional calculation of PrFeAsO in NdFeAsO structure shows that it
has the same orbital features as NdFeAsO, indicating that the major 
role of different Ln atoms is to change structural parameters. 
Thus the Ln-dependent gradual change of the total energy of $d_{yz}$- 
and $d_{xy}$-AFMs, shown in Fig.~2, is mainly from the change in  
structural parameters including the As height from the Fe plane. 
Considering overall increase of measured $T_c$ in doped LnFeAsO from La to 
Gd,\cite{kamihara1,chenxh,chengf,ren1,ren2,miyazawa,kamihara2,wang}
the different Ln-dependences of the calculated total energy in 
the $d_{yz}$- and $d_{xy}$-AFMs suggest that electrons in Fe~$d_{yz}$ and 
$d_{xy}$ orbitals may play different roles in SC.


\begin{figure} 
\epsfig{file=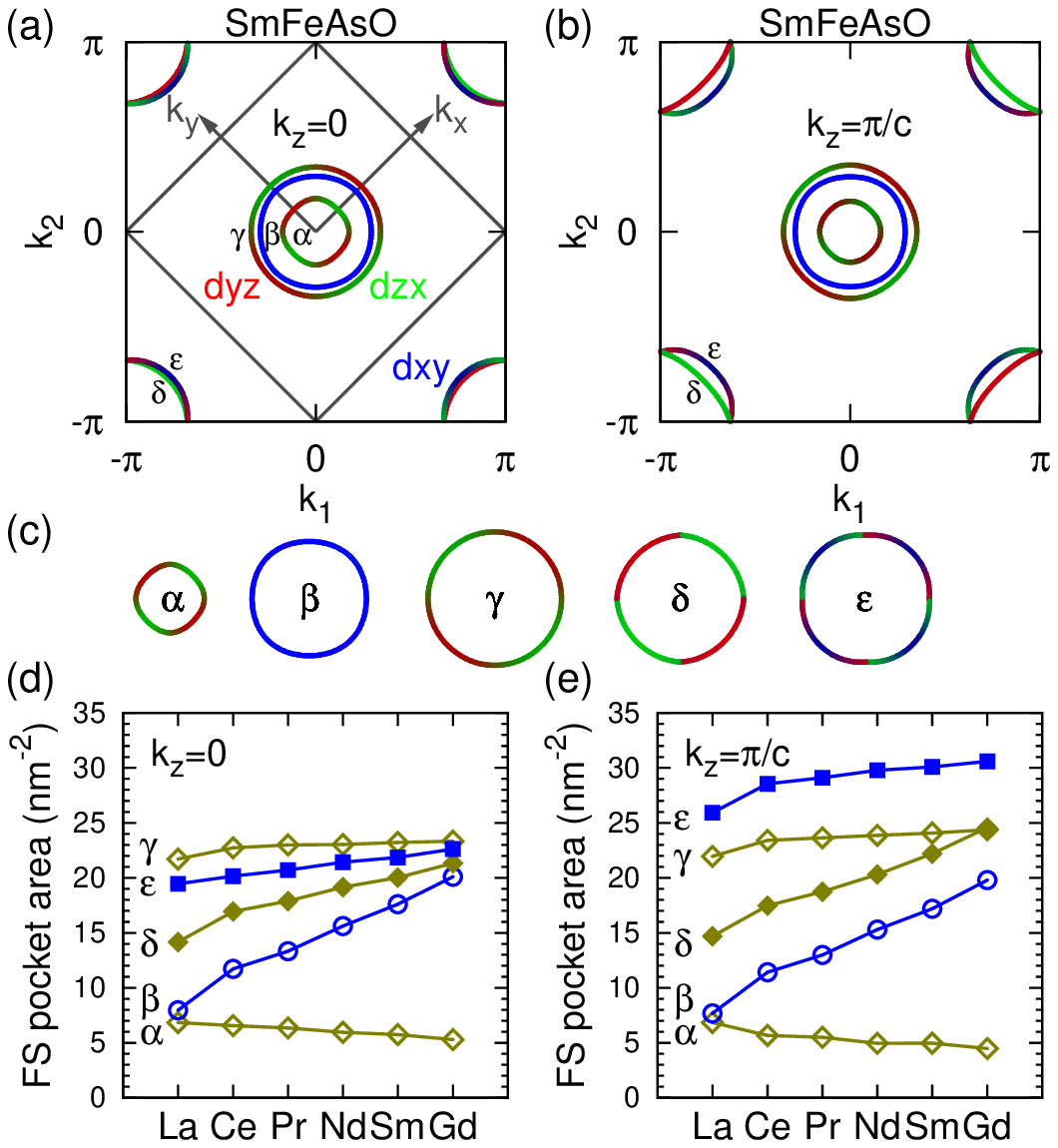,width=8.5cm,angle=0.0,clip=} 
\caption{(Color online) Orbital characters and pocket areas of Fermi 
surfaces (FSs) in nonmagnetic LnFeAsO. (a) FS of nonmagnetic SmFeAsO 
at $k_z$ = 0 and (b) at $k_z$ = $\pi/c$. The $k_1$ and $k_2$ axes are along 
reciprocal lattice vectors of the nonmagnetic unit cell having two Fe 
atoms. In (a), the 45$^\circ$-rotated inner square is the first 
Brillouin zone of the $\sqrt{2}\times\sqrt{2}\times1$ supercell having 
four Fe atoms, where $k_x$ and $k_y$ axes are along the $x$ and $y$ axes in 
Fig.~1(a), respectively. Three hole pockets ($\alpha$, $\beta$, and 
$\gamma$) are located at the center and two electron pockets ($\delta$ and
$\epsilon$) at the corner. Orbital characters, $d_{xy}$, $d_{yz}$, 
and $d_{zx}$, are represented in blue, red, and green, respectively. 
The $\alpha$, $\gamma$, and $\delta$ pockets are derived from $d_{yz}$ and 
$d_{zx}$, the $\beta$ pocket is from $d_{xy}$, and the $\epsilon$ 
pocket is from $d_{xy}$, $d_{yz}$, and $d_{zx}$.  (c) FS pockets at $k_z$ = 0. 
The $\delta$ and $\epsilon$ pockets are plotted as closed loops to highlight 
changes in orbital characters around the pockets. (d) Areas of the five
FS pockets at $k_z$ = 0 and (e) at $k_z$ = $\pi/c$ as a function of Ln 
elements.
}
\end{figure}

To understand the origin of the distinctive orbital characters of the Fe 
magnetic moments, we analyzed orbital characters of electronic states at 
the Fermi energy in nonmagnetic LnFeAsO. As a prototype, SmFeAsO has 
three cylindrical hole-type FS pockets [$\alpha$, $\beta$, and $\gamma$ 
pockets in Figs.~3(a)-(c)] and two electron-type FS pockets [$\delta$ and 
$\epsilon$ pockets in Figs.~3(a)-(c)] at $E_F$. Orbital analysis shows that 
the $\alpha$, $\gamma$, and $\delta$ pockets are derived from Fe $d_{yz}$ 
and $d_{zx}$ orbitals, the $\beta$ pocket is from $d_{xy}$, and the 
$\epsilon$ pocket is from $d_{xy}$, $d_{yz}$, and $d_{zx}$ [Figs.~3(a)-(c)]. 
For FS nesting, we consider six pairs of hole and electron pockets: 
$\alpha$-$\delta$, $\beta$-$\delta$, $\gamma$-$\delta$, $\alpha$-$\epsilon$, 
$\beta$-$\epsilon$, and $\gamma$-$\epsilon$ pairs. Among these, 
$\alpha$-$\delta$ nesting is not significant because the two pockets have 
quite different pocket areas in all LnFeAsO [Figs.~3(d) and (e)], and 
$\beta$-$\delta$ nesting is not effective because orbital difference of the 
two pockets degrades effects of nesting greatly. In addition, 
$\gamma$-$\delta$ and $\alpha$-$\epsilon$ nesting effects should be very 
weak because of orbital mismatch due to out-of-phase alternations of 
$d_{yz}$ and $d_{zx}$ characters around the pockets. Thus, we need to 
examine only $\beta$-$\epsilon$ and $\gamma$-$\epsilon$ nesting.

As mentioned above, the $\beta$-hole pocket in LnFeAsO [Figs.~3(a)-(c)] 
is derived from the $d_{xy}$ orbital, the $\gamma$-hole pocket is from
$d_{yz}$ and $d_{zx}$ orbitals, and the $\epsilon$-electron pocket is 
from $d_{xy}$, $d_{yz}$, and $d_{zx}$ orbitals. Thus we need to analyze 
whether the $\beta$-$\epsilon$ nesting is related with $d_{xy}$ orbital 
character of the Fe magnetic moment and whether the $\gamma$-$\epsilon$ 
nesting is related with $d_{yz}$ or $d_{zx}$ orbital character of the 
moment. Calculated FS pocket areas in LnFeAsO [Figs.~3(d) and (e)] show 
that the $\beta$-hole pocket area depends sensitively on Ln elements and 
it is closer to the $\epsilon$-electron pocket area for heavier Ln elements, 
implying that $\beta$-$\epsilon$ nesting is better for heavier Ln elements. 
Considering the increase of stability of $d_{xy}$-AFM from La to Gd 
[Fig.~2], we find that the FS nesting between the $\beta$ and $\epsilon$ 
pockets in LnFeAsO has positive correlation with the $d_{xy}$ orbital 
character of the Fe magnetic moment. On the contrary, the $\gamma$-hole 
pocket area is almost insensitive to Ln elements and close to the 
$\epsilon$-electron pocket area [Figs.~3(d) and (e)], indicating that 
the $\gamma$-$\epsilon$ nesting is good for all LnFeAsO. Considering the 
decrease of stability of $d_{yz}$-AFM from La to Gd [Fig.~2], we conclude 
that the FS nesting between the $\gamma$ and $\epsilon$ pockets in 
LnFeAsO is hardly related with the $d_{yz}$ orbital character of the Fe 
magnetic moment.

Although it is not involved in the FS nesting, the area of the 
$\delta$-electron FS pocket, which is derived from $d_{yz}$ and $d_{zx}$ 
orbitals, also depends sensitively on Ln elements [Figs.~3(d) and (e)]. 
With heavier Ln elements, the area of the $\delta$-electron pocket increases 
and so does the area of the $\beta$-hole pocket which is derived from the 
$d_{xy}$ orbital.\cite{F_doping} This results in electron transfer from 
the $d_{xy}$ orbital to the $d_{yz}$ and $d_{zx}$ orbitals. Since the 
$\delta$-pocket is not related with FS nesting, the increased $d_{yz}$ and 
$d_{zx}$ electrons in the $\delta$-pocket can contribute to AFM only by 
Heisenberg-type local-moment interactions.\cite{yildirim,moon}
The overall decrease of stability of $d_{yz}$-AFM from LaFeAsO to GdFeAsO 
can be understood by the superexchange mechanism. For $d_{yz}$ orbitals of 
nearest-neighboring Fe atoms, it is known that indirect hopping through 
As $p$ orbitals is larger than direct hopping.\cite{calderon,haule2,yin4} 
From La to Gd, the As height from the Fe plane increases, so the Fe-As-Fe 
angle decreases, reducing the indirect hopping and thereby weakening the 
strength of Fe magnetic moment of the $d_{yz}$ orbital character. Thus, 
the stability of AFM from the Fe $d_{yz}$ orbital is consistent with the 
Heisenberg-type local-moment interaction due to the superexchange mechanism.


\begin{figure} 
\epsfig{file=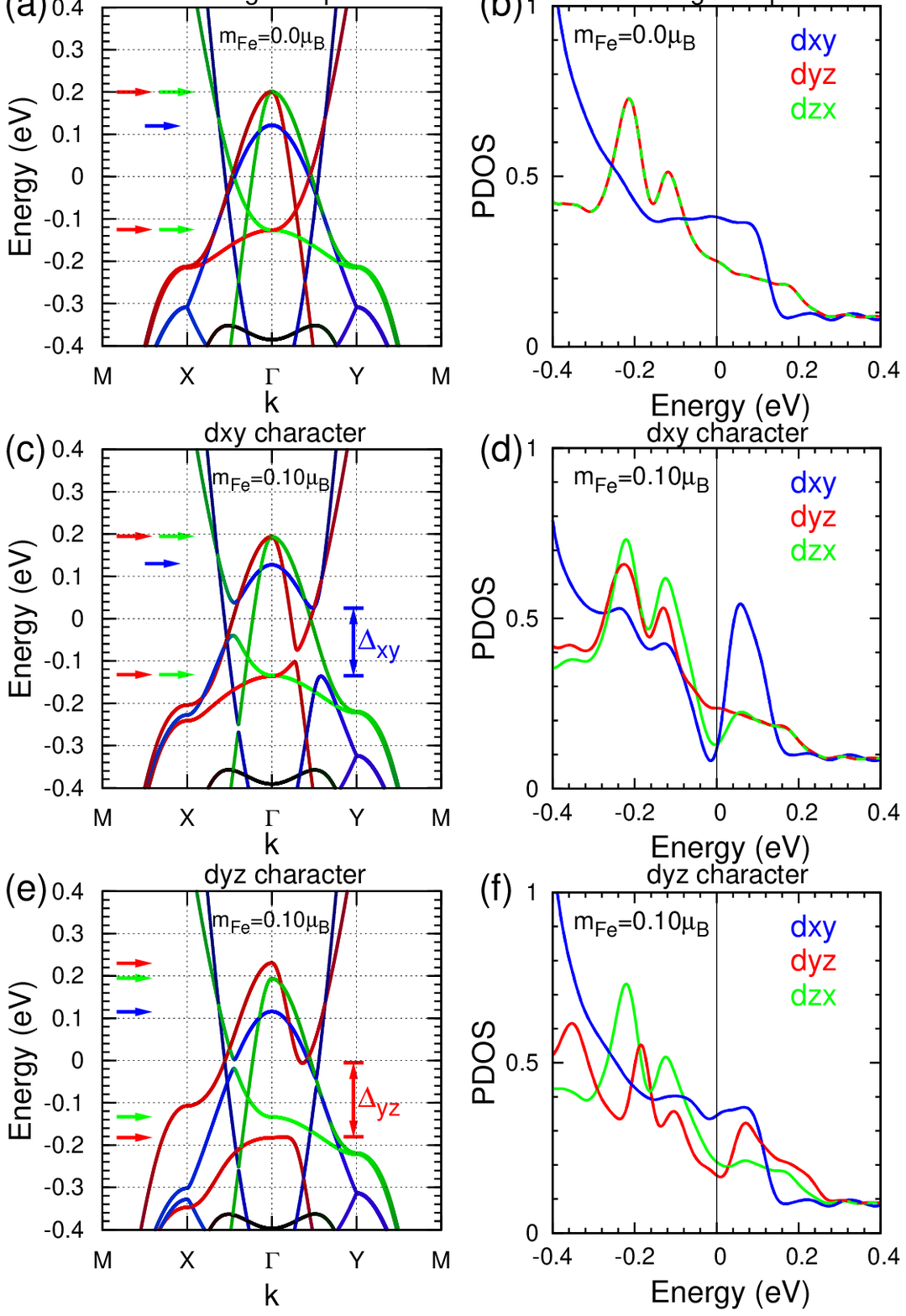,width=8.5cm,angle=0,clip=} 
\caption{(Color online) Orbital-resolved electronic band structures and 
projected density of state (PDOS) in SmFeAsO: (a) and (b) for nonmagnetic 
phase, (c) and (d) for the $d_{xy}$-AFM with $m_\text{Fe}$ = 
0.10~$\mu_\text{B}$, and (e) and (f) for the $d_{yz}$-AFM with 
$m_\text{Fe}$ = 0.10~$\mu_\text{B}$. Band structures are plotted along 
high-symmetry lines in the first Brillouin zone of the 
$\sqrt{2}\times\sqrt{2}\times1$ supercell for the 
single-stripe-type AFM, where M = (0, $\pi$), X = ($\pi$/2, $\pi$/2), 
$\Gamma$ = (0, 0), and Y = (-$\pi$/2, $\pi$/2) with respect to the unit-cell 
reciprocal axes, $k_1$ and $k_2$, shown in Fig. 3(a). Blue, red, and green 
represent $d_{xy}$, $d_{yz}$, and $d_{zx}$ orbital characters, respectively, 
and $E_F$ is set to zero. Horizontal arrows indicate extreme band energies 
at $\Gamma$. In (c), $\Delta_{xy}$ indicates energy gap between $d_{xy}$
bands in the $\Gamma$Y line. In (e), $\Delta_{yz}$ indicates energy gap 
between $d_{yz}$ bands in the $\Gamma$Y line.
}
\end{figure}

To find out effects of different orbital characters of Fe magnetic moments on 
the electronic structures, we obtained orbital-resolved band structures and 
projected density of states (PDOS) in SmFeAsO, as a prototype of LnFeAsO, 
in nonmagnetic phase and $d_{xy}$- and $d_{yz}$-AFMs (Fig.~4). In the 
nonmagnetic phase with zero $m_\text{Fe}$, energy bands have distinct 
orbital characters along high-symmetry lines, yielding PDOS slowly varying 
near $E_F$ [Figs.~4(a) and (b)].

In SmFeAsO in $d_{xy}$-AFM where Fe magnetic moment has only $d_{xy}$ orbital 
character, strong anti-crossing occurs between $d_{xy}$ bands, opening 
orbital-dependent energy gaps at $E_F$ [Fig.~4(c)]. In the $\Gamma$X line, 
$d_{xy}$ and $d_{zx}$ bands repel each other at $E_F$ and at -260~meV. In 
the $\Gamma$Y line, $d_{xy}$ bands open a significant energy gap near 
$E_F$ [marked with $\Delta_{xy}$ in Fig.~4(c)], while $d_{zx}$ bands are 
intact. Except for anti-crossing at -100~meV in the $\Gamma$Y line, 
$d_{yz}$ bands are almost unchanged. The $d_{xy}$ PDOS is greatly modified 
with a partial-gap opening at $E_F$ and a huge peak right above $E_F$ 
[Fig.~4(d)]. The $d_{zx}$ PDOS also shows a reduction near $E_F$ because 
of coupling to $d_{xy}$ bands [Fig.~4(d)]. Despite these significant 
effects on the electronic structures near $E_F$, deformation of bands is 
mostly confined near the band-crossing points, with no shift of original 
band edges [Fig.~4(c)]. This confirms that the Fe magnetic moment of 
$d_{xy}$ orbital character is due to FS nesting.

In contrast, in SmFeAsO in $d_{yz}$-AFM where the Fe magnetic moment has only 
$d_{yz}$ orbital character, $d_{yz}$ bands are deformed in the whole 
Brillouin zone rather than just anti-crossing of bands [Fig.~4(e)]. Whole 
upper part of $d_{yz}$ bands and whole lower part of $d_{yz}$ bands are 
repelled from each other [Fig.~4(e)], so even the top of the hole-type
$d_{yz}$ band is pushed up by 40~meV and the bottom of the electron-type 
$d_{yz}$ band is pushed down by 50~meV. This change of $d_{yz}$ bands 
in the whole Brillouin zone indicates that the driving mechanism for the 
$d_{yz}$ orbital character of Fe magnetic moment is not FS nesting which
is local in k-space but it is the lowering of the total energy of the 
electron system by formation of magnetic moments which are local in real 
space. Despite the significant change in the $d_{yz}$ bands, 
all the other bands are almost unchanged except for some anti-crossings.

As shown in Fig.~4, Fe magnetic moments with different orbital characters 
couple differently to electrons at the Fermi energy, so they may have 
different roles in SC and AFM. In particular, detailed comparison of our 
band structures with experimental results, e.g., angle-resolved 
photoemission spectroscopy of detwinned samples,\cite{yi,kim} will reveal 
orbital characters of the Fe magnetic moment, and thereby their roles in AFM.
In addition, when $m_\text{Fe}$ is increased to 1.0 $\mu_\text{B}$ or 
larger, the Fe magnetic moment evolves gradually to an almost spherical 
shape with no specific orbital characters. Thus, the orbital-distinctive 
magnetic moments and their effects will appear with different strengths in 
various iron pnictides and chalcogenides depending on the wide range
of reported $m_\text{Fe}$.

In conclusion, we have found that Fe magnetic moment in LnFeAsO (Ln = La 
to Gd) has distinctive orbital character, $d_{xy}$ or $d_{yz}$, when the
Fe magnetic moment is small and ordered antiferromagnetically along the
$x$ direction and ferromagnetically along the $y$ direction. By comparing 
atomic structures and calculated $d$-orbital-resolved electronic structures 
of the compounds,
we have shown that the origins of the $d_{xy}$ and $d_{yz}$ orbital 
characters of the Fe magnetic moment are orbital-dependent FS nesting and 
superexchange interactions, respectively. Fe magnetic moments of $d_{xy}$ 
and $d_{yz}$ orbital characters are found to have different coupling strengths
to electronic states near the Fermi energy, so they can be identified by 
$d$-orbital-resolved measurement of electronic structures and they may have 
different roles in SC and AFM.

This work was supported by the NRF of Korea (Grant Nos. 2009-0081204 
and 2011-0018306). Computational resources have been provided by KISTI 
Supercomputing Center (Project No. KSC-2008-S02-0004).


\begin{thebibliography}{99}

\bibitem{kamihara1}
Y. Kamihara, T. Watanabe, M. Hirano, and H. Hosono, J. Am. Chem. Soc. {\bf 130}, 3296 (2008).

\bibitem{chenxh}
X. H. Chen, T. Wu, G. Wu, R. H. Liu, H. Chen, and D. F. Fang, Nature (London) {\bf 453}, 761 (2008).

\bibitem{cruz}
C. de la Cruz, Q. Huang, J. W. Lynn, J. Li, W. Ratcliff II, J. L. Zarestky, H. A. Mook, G. F. Chen, J. L. Luo, N. L. Wang, and P. Dai, Nature (London) {\bf 453}, 899 (2008).

\bibitem{zhao1}
J. Zhao, Q. Huang, C. de la Cruz, S. Li, J. W. Lynn, Y. Chen, M. A. Green, G. F. Chen, G. Li, Z. Li, J. L. Luo, N. L. Wang, and P. Dai, Nature Materials {\bf 7}, 953 (2008).

\bibitem{yin}
Z. P. Yin, S. Leb\`egue, M. J. Han, B. P. Neal, S. Y. Savrasov, and W. E. Pickett, Phys. Rev. Lett. {\bf 101}, 047001 (2008).

\bibitem{mazin}
I. I. Mazin, D. J. Singh, M. D. Johannes, and M. H. Du,
Phys. Rev. Lett. {\bf 101}, 057003 (2008).

\bibitem{yildirim}
T. Yildirim, Phys. Rev. Lett. {\bf 101}, 057010 (2008).

\bibitem{chengf}
G. F. Chen, Z. Li, D. Wu, G. Li, W. Z. Hu, J. Dong, P. Zheng, J. L. Luo, and N. L. Wang, Phys. Rev. Lett. {\bf 100}, 247002 (2008).

\bibitem{qiu}
Y. Qiu, W. Bao, Q. Huang, T. Yildirim, J. M. Simmons, M. A. Green, J. W. Lynn, Y. C. Gasparovic, J. Li, T. Wu, G. Wu, and X. H. Chen, Phys. Rev. Lett. {\bf 101}, 257002 (2008).

\bibitem{zhao2}
J. Zhao, Q. Huang, C. de la Cruz, J. W. Lynn, M. D. Lumsden, Z. A. Ren, J. Yang, X. Shen, X. Dong, Z. Zhao, and P. Dai, Phys. Rev. B {\bf 78}, 132504 (2008).

\bibitem{ren1}
Z. A. Ren, J. Yang, W. Lu, W. Yi, G. C. Che, X. L. Dong, L. L. Sun, and Z. X. Zhao, Mater. Res. Innov. {\bf 12}, 105 (2008).

\bibitem{ren2}
Z. Ren, W. Lu, J. Yang, W. Yi, X. Shen, Z. Li, G. Che, X. Dong, L. Sun, F. Zhou, and Z. Zhao, Chin. Phys. Lett. {\bf 25}, 2215 (2008).

\bibitem{miyazawa}
K. Miyazawa, K. Kihou, P. M. Shirage, C. Lee, H. Kito, H. Eisaki, and A. Iyo, J. Phys. Soc. Jpn. {\bf 78}, 034712 (2009).

\bibitem{kamihara2}
Y. Kamihara, T. Nomura, M. Hirano, J. E. Kim, K. Kato, M. Takata, Y. Kobayashi, S. Kitao, S. Higashitaniguchi, Y. Yoda, M. Seto, and H. Hosono, New J. Phys. {\bf 12}, 033005 (2010).

\bibitem{wang}
P. Wang, Z. M. Stadnik, C. Wang, G. Cao, and Z. Xu, J. Phys. Condens. Matter {\bf 22}, 145701 (2010).

\bibitem{moon}
C. Y. Moon and H. J. Choi, Phys. Rev. Lett. {\bf 104}, 057003 (2010).

\bibitem{kuroki}
K. Kuroki, H. Usui, S. Onari, R. Arita, and H. Aoki, Phys. Rev. B {\bf 79}, 
224511 (2009).

\bibitem{lebegue}
S. Leb\`egue, Phys. Rev. B {\bf 75}, 035110 (2007).

\bibitem{lu}
D. H. Lu, M. Yi, S. K. Mo, A. S. Erickson, J. Analytis, J. H. Chu, D. J. Singh, Z. Hussain, T. H. Geballe, I. R. Fisher, and Z. X. Shen,
Nature (London) {\bf 455}, 81 (2008).

\bibitem{singh}
D. J. Singh and M. H. Du,
Phys. Rev. Lett. {\bf 100}, 237003 (2008).

\bibitem{haule}
K. Haule, J. H. Shim, and G. Kotliar,
Phys. Rev. Lett. {\bf 100}, 226402 (2008).

\bibitem{moon_2008} 
C.-Y. Moon, S. Y. Park, and H. J. Choi, 
Phys. Rev. B {\bf 78}, 212507 (2008).

\bibitem{lee}
C.-C. Lee, W.-G. Yin, and W. Ku,
Phys. Rev. Lett. {\bf 103}, 267001 (2009).

\bibitem{kruger}
F. Kr\"uger, S. Kumar, J. Zaanen, and J. van den Brink, 
Phys. Rev. B {\bf 79}, 054504 (2009).

\bibitem{moon_2009} 
C.-Y. Moon, S. Y. Park, and H. J. Choi, 
Phys. Rev. B {\bf 80}, 054522 (2009).

\bibitem{chencc}
C. C. Chen, J. Maciejko, A. P. Sorini, B. Moritz, R. R. P. Singh, and, T. P. Devereaux, Phys. Rev. B {\bf 82}, 100504(R) (2010).

\bibitem{lv}
W. Lv, F. Kr\"uger, and P. Phillips,
Phys. Rev. B {\bf 82}, 045125 (2010).

\bibitem{yi}
M. Yi, D. Lu, J. Chu, J. G. Analytis, A. P. Sorini, A. F. Kemper, B. Moritz, S. Mo, R. G. Moore, M. Hashimoto, W. Lee, Z. Hussain, T. P. Devereaux, I. R. Fisher, and Z. X. Shen, Proc. Natl. Acad. Sci. USA {\bf 108}, 6878 (2011).

\bibitem{lv2}
W. Lv, J. Wu, and P. Phillips, Phys. Rev. B {\bf 80}, 224506 (2009).

\bibitem{yin2}
Z. P. Yin, K. Haule, and G. Kotliar, Nature Physics {\bf 7}, 294 (2011).

\bibitem{daghofer}
M. Daghofer, Q.-L. Luo, R. Yu, D. X. Yao, A. Moreo, and E. Dagotto, Phys. Rev. B {\bf 81}, 180514(R) (2010).

\bibitem{yin3}
Z. P. Yin and W. E. Pickett, Phys. Rev. B {\bf 81}, 174534 (2010).

\bibitem{sanchez}
D. S$\Acute{a}$nchez-Portal, P. Ordej\'on, E. Artacho, and J. M. Soler,
Int. J. Quantum Chem. {\bf 65}, 453 (1997).

\bibitem{nekrasov}
I. A. Nekrasov, Z. V. Pchelkina, and M. V. Sadovskii,
JETP Letters {\bf 87}, 560 (2008).

\bibitem{pourovskii}
L. Pourovskii, V. Vildosola, S. Biermann, and A. Georges, Europhys. Lett. {\bf 84}, 37006 (2008).

\bibitem{martinelli}
A. Martinelli, A. Palenzona, C. Ferdeghini, M. Putti, and H. Emerich, J. Alloys Comp. {\bf 477}, L21 (2009).

\bibitem{structure_optimization}
While our spin-unpolarized calculation predicts the lattice constants 
of the nonmagnetic tetragonal phase accurately, it underestimates 
the As height by 0.1 {\AA} because it neglects non-zero Fe magnetic 
moment which is actually present even at high-temperature nonmagnetic 
phase. The As height is predicted correctly by our spin-polarized 
calculation. 

\bibitem{T_vs_O}
Our test calculations show that using an orthorhombic unit cell 
rather than a tetragonal one produces only a small change in 
orbital occupation of $d_{xz}$ and $d_{yz}$, with no significant 
change in our main results of $d_{xy}$ versus $d_{yz}$ orbital characters.

\bibitem{F_doping}
We also calculated Fermi-surface pocket areas in LnFeAsO$_{0.9}$F$_{0.1}$ 
using the virtual crystal approximation, and found that the F doping does 
not alter Ln-dependent changes of $\beta$- and $\delta$-pocket areas 
although the F doping shrinks hole-type pockets and enlarges electron-type 
pockets compared with undoped cases.

\bibitem{calderon}
M. J. Calder\'on, B. Valenzuela, and E. Bascones, 
Phys. Rev. B {\bf 80}, 094531 (2009).

\bibitem{haule2}
K. Haule and G. Kotliar, New J. Phys. {\bf 11}, 025021 (2009).

\bibitem{yin4}
Z. P. Yin, K. Haule, and G. Kotliar, 
Nature Materials {\bf 10}, 932 (2011).

\bibitem{kim}
Y. Kim, H. Oh, C. Kim, D. Song, W. Jung, B. Kim, H. J. Choi, C. Kim, B. Lee, S. Khim, H. Kim, K. Kim, J. Hong, and Y. Kwon, Phys. Rev. B {\bf 83}, 064509 (2011).


\end{thebibliography}
\end{document}